\documentclass[12pt]{iopartCN}

\usepackage{graphicx}
\begin{document}

\title[Xiong et al.: Response of the Earth's magnetosphere and ionosphere]
{Response of the Earth's magnetosphere and ionosphere to solar
wind driver and ionosphere load: Results of global MHD
simulations\footnote[7]{Corresponding Author: Xiong Ming
(mxiong@ustc.edu.cn)
\\
Supported by the National Natural Science Foundation of China
(40831060, 40621003, 40774077), the China Postdoctoral Science
Foundation (20070420725) and K.C.Wong Education Foundation of Hong
Kong.}}

\author{XIONG Ming, PENG Zhong, HU Youqiu,
ZHENG Huinan}

\address{CAS Key Laboratory for Basic Plasma Physics,
School of Earth and Space Sciences,
University of Science and Technology of China, Hefei Anhui 230026, China}

\begin{abstract}
Three-dimensional global magnetohydrodynamic simulations of the
solar wind - magnetosphere - ionosphere system are carried out to
explore the dependence of the magnetospheric reconnection voltage,
the ionospheric transpolar potential, and the field aligned
currents (FACs) on the solar wind driver and ionosphere load for
the cases with pure southward interplanetary magnetic field (IMF).
It is shown that the reconnection voltage and the transpolar
potential increase monotonically with decreasing Pedersen
conductance ($\Sigma_{\rm P}$), increasing southward IMF strength
($B_{\rm s}$) and solar wind speed ($v_{\rm sw}$). Moreover, both
of the region 1 and the region 2 FACs increase when $B_{\rm s}$
and $v_{\rm sw}$ increase, whereas the two currents behave
differently in response to $\Sigma_{\rm P}$. As $\Sigma_{\rm P}$
increases, the region 1 FAC increases monotonically, but the
region 2 FAC shows a non-monotonic response to the increase of
$\Sigma_{\rm P}$: it first increases in the range of (0, 5)
Siemens and then decreases for $\Sigma_P >$ 5 Siemens.

\end{abstract}


\submitto{Chinese Physics Letters}

\vspace{\baselineskip}


The solar wind - magnetosphere - ionosphere (SW-M-I) system is a
multi-variable, multi-scale complex system \cite{Raeder2003}, in
which the solar wind serves as a driver and the ionosphere plays a
role of load. It has been widely believed that magnetic
reconnection (MR) between the geomagnetic and interplanetary
fields on the dayside magnetopause leads to an efficient and rapid
transfer of energy and momentum from the solar wind to the
magnetosphere. The electrons may be temporarily trapped in the
central reconnection region including electron diffusion region
\cite{He2008}. The MR takes place along the separatrix line, i.e.,
the MR line, which connects a pair of magnetic nulls lying on the
magnetopause \cite{Priest2000}. The total electric potential drop
along the MR line (MR voltage) virtually represents the global MR
rate \cite{Siscoe2001,Hu2007} so as to be an important parameter
characterizing the SW-M-I coupling. The ionospheric transpolar
potential, named as ``cross polar cap potential" sometimes in the
literature and defined as the difference between the positive and
negative potential peaks in the ionosphere, was taken by some
authors as a measure for the MR voltage \cite{Fedder1995}, but
actually they differ from each other because of the presence of
parallel electric field along the geomagnetic field lines passing
through the reconnection region \cite{Siscoe2001,Hu2007}. After
all, the transpolar potential remains to be another coupling
parameter, which is presently observable.

The significant progress has been made in the high latitude
boundary/cusp region by the Cluster observation, such as the
triple cusps due to temporal or spatial effect \cite{Zong2004}, a
plasmoid-like structure without a core magnetic field
\cite{Zong2005a}, association of reverse convection with cusp
proton aurora \cite{Zong2005b}. The high-latitude ionosphere
serves as a monitoring screen for magnetospheric activities due to
the magnetosphere-ionosphere (M-I) coupling. The coupling consists
of a mapping of field aligned currents (FACs) from the
magnetosphere to the ionosphere and a feedback of the ionospheric
potential to the magnetosphere, which drives the magnetospheric
plasma convection. The FAC generally consists of region 1 current
on the poleward side and region 2 current on the equatorward side,
and the two FAC currents flow in opposite sense \cite{Iijima1976}.
The so-named northern $B_z$ (NBZ) current appears when the IMF is
northward \cite{Song1999}, but it does not exist on the condition
of southward IMF \cite{Tanaka1995}. A direct manifestation of the
FACs is the aurora within the polar cap, so one may infer the
global magnetospheric process from the morphology, intensity, and
duration of the aurora. Therefore, a great attention has been paid
to the investigation of the FACs.

The M-I system is jointly decided by various parameters such as
the interplanetary magnetic field (IMF), the solar wind ram
pressure, and the ionospheric conductance. When any of these
parameters changes, the M-I system will accommodate itself and
reach a new equilibrium. This results in a change of the MR
voltage, the transpolar potential, and the FAC intensity. Note
that here the ionosphere is not only a passive load for the M-I
system, but plays a crucial role in the M-I coupling. It directly
affects the transpolar potential and the FAC intensity, and exerts
an indirect but important influence on the magnetosphere
configuration, the magnetosheath flow, and the MR voltage
\cite{Merkin2003, Merkin2005}. Hence the effect of the ionospheric
conductance must be included while studying the SW-M-I coupling.
The governing equations and initial boundary conditions for this
MHD model were given in detail by Hu et al. \cite{Hu2007}.

Global MHD simulation is a powerful tool for the study of the
response of the M-I system to the solar wind driver and ionosphere
load. As a matter of fact, it was used to study the dependence of
the MR voltage and transpolar potential on the solar wind electric
field and ionospheric Pedersen conductance
\cite{Siscoe2001,Hu2007,Merkin2003,Merkin2005} and the paths of
the FACs \cite{Tanaka1995, Janhunen1997, Guo2008}. In particular,
Guo et al. \cite{Guo2008} found that more than 50\% of the region
1 FAC may come from the bow shock under strong southward IMF
conditions. To our knowledge, however, a systematic study of the
dependence of the MR voltage, transpolar potential, and FAC
intensity on the solar wind driver and ionospheric conductance is
unavailable in the literature, and this paper will undertake such
a task. All simulations in this study are made with the use of the
PPMLR-MHD numerical scheme developed by Hu et al.
\cite{Hu2007,Hu2005}.

\begin{table} 
\vspace*{2cm}%
\caption{Simulation case assortment of parametric studies of the
ionospheric conductance $\Sigma_{\rm P}$ and IMF $B_{\rm s}$}\label{Tab1}
\begin{tabular}{c|c|c|c|c}
\hline
Group & Case & $v_{\rm sw}$ (km/s) & $B_{\rm s}$ (nT) & $\Sigma_{\rm P}$ (S) \\[0pt]
\hline  
1 & A$_1$, B$_1$, C$_1$, D$_1$, E$_1$ & 400 & 5 & 0.1, 1, 5, 10, 20\\[0pt]
\hline
2 & A$_2$, B$_2$, C$_2$, D$_2$, E$_2$ & 800 & 5 & 0.1, 1, 5, 10, 20\\[0pt]
\hline \hline
3 & F$_1$, G$_1$, H$_1$, I$_1$, J$_1$ & 400 & 2, 5, 10, 15, 20 & 5\\[0pt]
\hline
4 & F$_2$, G$_2$, H$_2$, I$_2$, J$_2$ & 800 & 2, 5, 10, 15, 20 & 5\\[0pt]
\hline
\end{tabular}
\end{table}

In order to reflect the fundamental physics of the SW-M-I system,
two simplified assumptions are made: (1) the IMF is pure
southward, and (2) the ionospheric Pedersen conductance is uniform
and the Hall conductance vanishes. Three adjustable parameters are
chosen in the following numerical examples: the solar wind speed
$v_{\rm sw}$ and the southward IMF $B_{\rm s}$, which describe the
solar wind driver, and the Pedersen conductance $\Sigma_{\rm P}$,
which characterizes the ionosphere load. All other solar wind
parameters are fixed, including the number density $n = 5$
cm$^{-3}$ and temperature $T=0.91 \times 10^5$ K. The
corresponding adjustable parameters of numerical examples are
listed in the Table \ref{Tab1}, being divided into four groups: 1
and 2 for different values of $\Sigma_{\rm P}$, and 3 and 4 for
different values of $B_{\rm s}$. Groups 1 and 2 differ in $v_{\rm
sw}$, so do Groups 3 and 4. The MR voltage, the transpolar
potential, and the total region 1 and 2 current intensities, $I_1$
and $I_2$, are calculated for each example in order to examine
their dependence on $\Sigma_{\rm P}$, $B_{\rm s}$, and $v_{\rm
sw}$. The calculation of these quantities are straightforward
except for the MR voltage, which is evaluated with the use of the
method proposed by Hu et al. \cite{Hu2007}.

\begin{figure}[ht]
\begin{center}
\includegraphics[width=0.93\textwidth]{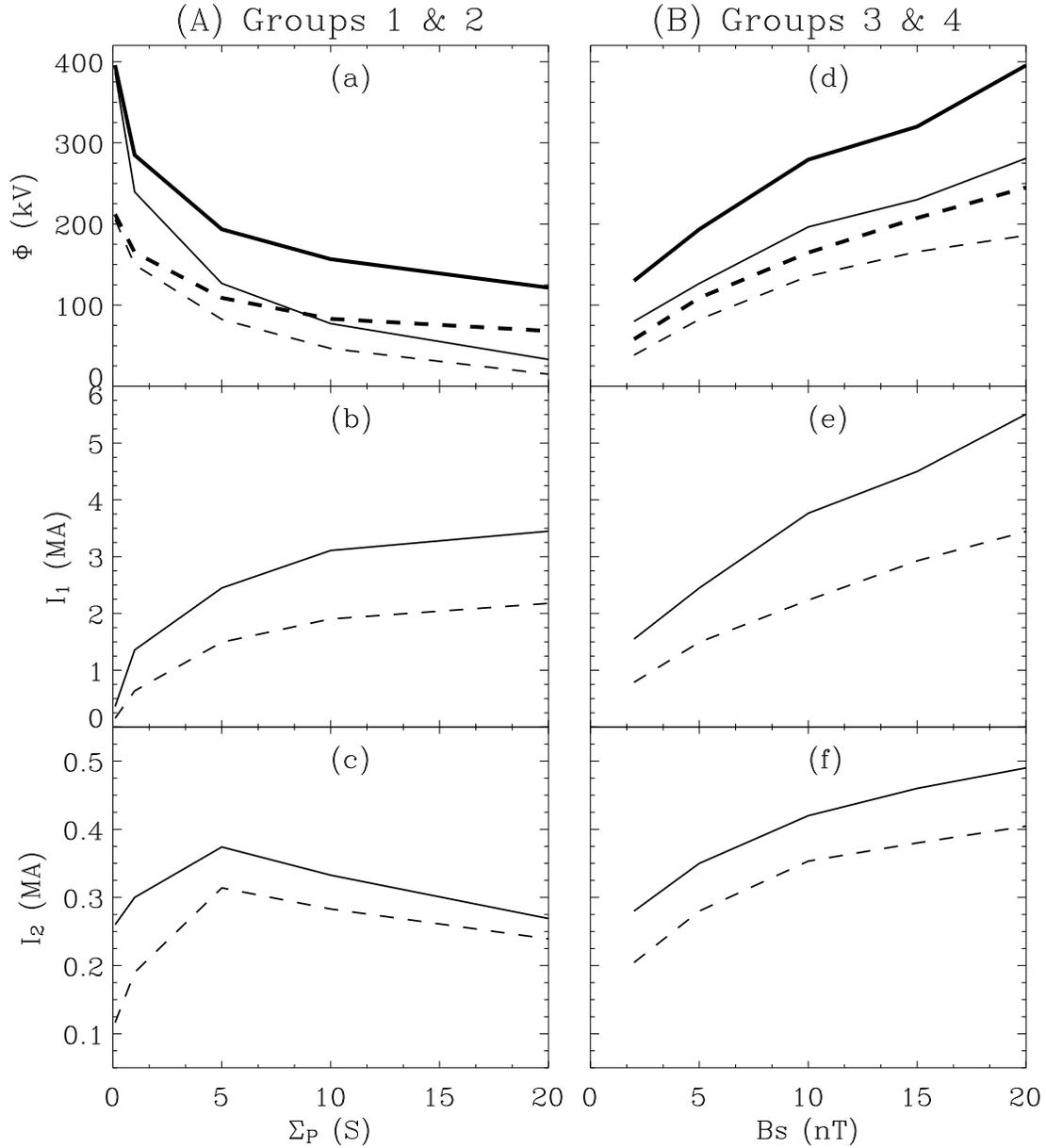}
\end{center}
\caption{\label{Fig:1} The responses of the MR voltage and
transpolar potential (a, d), region 1 FAC intensity $I_1$ (b, e),
and region 2 FAC intensity $I_2$ (c, f) as functions of the
Pedersen conductance $\Sigma_{\rm P}$ in column A (Groups 1 \& 2)
and the southward IMF $B_{\rm s}$ in column B (Groups 3 \& 4).
Numerical results are drawn in solid for Groups 1 and 3 with the
solar wind speed $v_{\rm sw}$ = 400 km$\cdot$s$^{-1}$ and in
dashed for Groups 2 and 4 with $v_{\rm sw}$ = 400
km$\cdot$s$^{-1}$. The thick and thin lines in (a, d) denote the
MR voltage and transpolar potential, respectively.}
\end{figure}

The responses of the MR voltage, transpolar potential, region 1
FAC $I_1$, and region 2 FAC $I_2$ to $\Sigma_{\rm P}$, $B_{\rm s}$
and $v_{\rm sw}$ are shown in Fig. 1. As seen from the left column
of Fig. 1, the MR voltage and the transpolar potential [Fig. 1(a)]
decrease monotonically with increasing $\Sigma_{\rm P}$. The MR
voltage is always larger than the transpolar potential and the two
become approximately identical for very small $\Sigma_{\rm P}$ or
$I_1$, the same conclusion as reached by Hu et al. \cite{Hu2007}.
The region 1 current intensity $I_1$ increases monotonically with
increasing $\Sigma_{\rm P}$ [Fig. 1(b)] as expected, whereas the
region 2 current $I_2$ exhibits a non-monotonic response to the
increase of $\Sigma_{\rm P}$ [Fig. 1(c)]: it first increases in
the range of (0, 5) S and then decreases as $\Sigma_{\rm P}$
exceeds 5 S. Two competing factors may be responsible for such a
behavior of $I_2$. On the one hand, the enhancement of
$\Sigma_{\rm P}$ favors an increase of $I_2$ as it does for $I_1$;
on the other hand, the weakening of the MR voltage caused by the
enhancement of $\Sigma_{\rm P}$ reduces the magnetospheric plasma
convection and the associated convectional electric field, which
in turn leads to a decrease of $I_2$. The former factor dominates
if $\Sigma_{\rm P}$ is small, whereas the latter becomes dominant
if $\Sigma_{\rm P}$ exceeds a certain critical value, 5 S for the
present cases. Therefore, $I_2$ exhibits a non-monotonic response
to $\Sigma_{\rm P}$ as illustrated in Fig. 1(c).

The response of the MR voltage, the transpolar potential, $I_1$
and $I_2$ to the southward IMF $B_{\rm s}$ and solar wind speed
$v_{\rm sw}$ is simpler: all of them increase monotonically when
$B_{\rm s}$ and $v_{\rm sw}$ increase. The two adjustable
parameters, $B_{\rm s}$ and $v_{\rm sw}$, characterize the solar
wind driver. The stronger the driver is, the larger the MR
voltage, the transpolar potential, and the region 1 and 2 FACs
will be.

From the results obtained we may reach the following conclusions.
(1) The ionosphere is not only a passive load, but also has a
subtle effect on the SW-M-I system. An increase of the ionospheric
Pedersen conductance $\Sigma_{\rm P}$ leads to a decrease of the
magnetospheric reconnection voltage and the ionospheric transpolar
potential, and an increase of the region 1 FAC. (2) The region 2
FAC shows a non-monotonic response to $\Sigma_{\rm P}$, consisting
an initial increase followed by a subsequent decrease with
increasing $\Sigma_{\rm P}$. Two competing factors, a positive
effect by the increase of $\Sigma_{\rm P}$ and a negative one by
the reduction of the MR voltage and the associated weakening of
the magnetospheric plasma convection, produce such a behavior of
the region 2 current. (3) All of the magnetospheric reconnection
voltage, the ionospheric transpolar potential, and the region 1
and 2 currents become larger under stronger solar wind conditions,
characterized by a larger solar wind ram pressure and a stronger
southward IMF.

Uniform Pedersen and zero Hall conductances are taken for the
ionosphere in this study for simplicity. In reality, the Hall
conductance does not vanish, and both of the Pedersen and Hall
conductances are non-uniform in space, fluctuating in time, and
closely related to the solar radiation intensities and solar wind
conditions \cite{Merkin2005}. While the present simulation model
needs to be refined to incorporate a more sophisticated treatment
of the ionospheric conductances, we argue that this will not
substantially change the basic conclusions reached above. Another
issue worth mentioning is that MHD simulations of the SW-M-I
coupling generally produce weak region 2 current
\cite{Song1999,Tanaka1995}. The contribution made by particle
drift to this current is important but essentially ignored in the
MHD simulations. Nevertheless, the MHD effect contributes to the
region 2 current, too, and this effect is properly reflected by
our preliminary simulations.

\end{document}